\documentclass[
 reprint,
 amsmath,amssymb,
 aip,
 apl
]{revtex4-2}

\usepackage{graphicx}
\usepackage{bm}
\usepackage{siunitx}
\usepackage{xcolor}
\usepackage{microtype}
\usepackage{placeins}
\usepackage{float}

\begin{document}

\title{Gatemon qubits based on in-situ ternary InAsSb/Al nanowire Josephson junctions}

\author{Zhenhai Sun}
\affiliation{NNF Quantum Computing Programme, Niels Bohr Institute, University of Copenhagen, \mbox{2100 Copenhagen, Denmark}}
\author{David Feldstein-Bofill}
\affiliation{NNF Quantum Computing Programme, Niels Bohr Institute, University of Copenhagen, \mbox{2100 Copenhagen, Denmark}}
\author{Sangeeth Kallatt}
\affiliation{NNF Quantum Computing Programme, Niels Bohr Institute, University of Copenhagen, \mbox{2100 Copenhagen, Denmark}}
\author{Sabbir A. Khan} 
\affiliation{Niels Bohr Institute, University of Copenhagen, 2100 Copenhagen, Denmark } 
\affiliation{Danish Fundamental Metrology, Kogle Alle 5, \mbox{2970 Hørsholm, Denmark}}
\author{Morten Kjaergaard}
\email[]{morten.kjaergaard@nbi.ku.dk}
\affiliation{NNF Quantum Computing Programme, Niels Bohr Institute, University of Copenhagen, \mbox{2100 Copenhagen, Denmark}}

\date{\today}

\begin{abstract}
Ternary InAsSb/Al nanowires combine compositionally engineerable semiconductor properties with epitaxial proximity superconductivity, making them attractive for hybrid quantum circuits and mesoscopic superconductivity.
Here, we realize gatemon qubits based on shadow-defined in-situ InAsSb/Al Josephson junctions in a single growth process.
Gate-dependent microwave spectroscopy demonstrates electrostatic tuning of the qubit transition over several gigahertz.
Rabi oscillations and Ramsey interference show coherent qubit control, with representative values $T_1=\SI{0.79}{\micro\second}$ and $T_2^*=\SI{0.91}{\micro\second}$. 
Across five gatemons distributed over three chips, the measured relaxation times range from approximately \SI{0.2}{\micro\second} to \SI{1.0}{\micro\second} and decrease with increasing qubit frequency. 
Their magnitude is comparable to a circuit estimate of spontaneous emission into the electromagnetic environment. 
These results establish in-situ InAsSb/Al junctions as coherent, gate-tunable circuit elements and introduce cQED architecture as a microwave probe of the dynamics of the ternary weak link.
\end{abstract}

\maketitle

Superconducting qubits conventionally employ superconductor--insulator--superconductor (SIS) tunnel junctions as nonlinear circuit elements~\cite{nakamura1999, krantz2019, kjaergaard2020}. 
A widely used superconducting qubit architecture is the transmon, where the SIS Josephson junction is shunted by a large capacitance~\cite{koch2007, barends2013}.
The transmon operates in the regime $E_J/E_C \gg 1$ to exponentially suppress offset-charge sensitivity, where $E_J$ is the Josephson energy and $E_C$ is the charging energy.
The resulted qubit frequency is given by $f_q = \sqrt{8E_JE_C}-E_C$.
A fixed-frequency transmon typically uses a single SIS junction, whereas frequency tunability is commonly achieved by replacing the junction with a superconducting quantum interference device (SQUID).
The effective Josephson energy $E_J(\Phi_\mathrm{ext})$, and thus the qubit frequency can then be controlled by an external magnetic flux $\Phi_\mathrm{ext}$.

\begin{figure}[!t]
\centering
\includegraphics[width=\columnwidth]{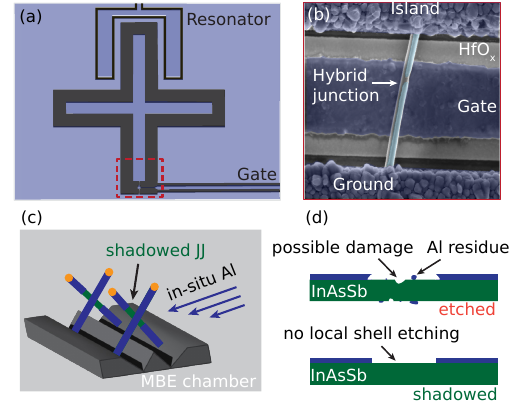}
\caption{Device architecture and formation of the shadow-defined junction. (a) Schematic layout of the capacitively shunted gatemon coupled to a microwave readout resonator. The dashed box indicates the junction region enlarged in panel (b). (b) Electron micrograph of the InAsSb nanowire junction between the superconducting island and ground. The shadow-defined weak link is located above the local gate and separated from it by an atomic-layer-deposited HfO$_x$ dielectric~\cite{khan2020}. (c) Schematic of directional in situ Al deposition. Geometrical shadowing leaves an uncovered semiconductor segment that forms the S--Sm--S weak link. (d) Schematic comparison between etched and shadow-defined junctions. In the etched approach, selective removal of the superconducting shell expose the weak link region to etchants, redisual Al, and surface damage. Shadow deposition defines the uncovered semiconductor segment witought local shell remocal in the active junction region.}
\label{fig:design}
\end{figure}

\begin{figure*}[t]
\centering
\includegraphics[width=\textwidth]{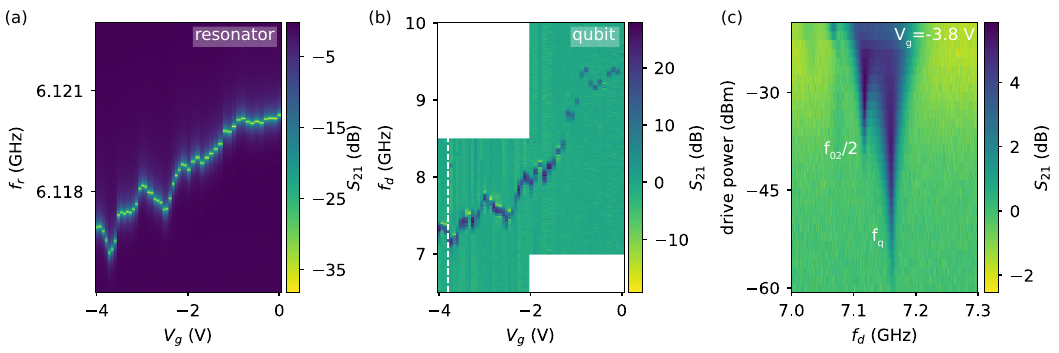}
\caption{Gate-dependent microwave spectroscopy of a shadow-junction gatemon. (a) Readout transmission magnitude as a function of readout frequency $f_r$ and junction-gate voltage $V_g$. (b) Two-tone spectroscopy versus drive frequency $f_d$ and $V_g$, showing electrical tuning of the qubit transition. The map combines two measurements with partially overlapping frequency windows; white regions correspond to frequencies that were not measured. The dashed line indicates the gate voltage selected for panel (c). (c) Power-dependent two-tone spectroscopy at $V_g=-3.8~\mathrm{V}$, showing broadening of the spectroscopic feature with increasing source power. In all panels, the color scale represents the transmission magnitude after subtraction of the mean response at each gate voltage or source power.}
\label{fig:gate_spectroscopy}
\end{figure*}

Replacing the insulating tunnel barrier with a semiconductor weak link (S--Sm--S) creates a gatemon, in which an electrostatic gate $V_g$ modifies the junction transmission and thereby controls the effective Josephson energy $E_J(V_g)$ and qubit frequency~\cite{aguado2020}. 
The semiconductor weak link also introduces microscopic degrees of freedom that are absent in a conventional tunnel junction. 
In particular, the Josephson potential depends on the number and transmission probabilities of the occupied Andreev levels~\cite{beenakker1991}, providing a connection between the qubit characteristic and the electronic structure of the weak link. 
Coherent gatemon operation has been demonstrated in several low-dimensional semiconductor platforms, including hybrid nanowires~\cite{larsen2015, delange2015, casparis2016, stavenga2023lower}, proximitized two-dimensional electron gases (2DEGs)~\cite{casparis2018, strickland2024}, selective-area-grown superconductor--semiconductor structures~\cite{hertel2022}, Germanium (Ge) heterostructures~\cite{sagi2024, kiyooka2025}, graphene~\cite{wang2019}, and carbon nanotubes~\cite{riechert2025}. 
Together, these experiments demonstrate that circuit quantum electrodynamics (cQED) can be extended beyond conventional tunnel junctions to investigate a broad range of hybrid materials and mesoscopic Josephson elements. 
In this setting, the semiconductor weak link provides an electrostatically tunable Josephson element, while the gatemon provides a sensitive cQED probe of the weak link and its underlying mesoscopic degrees of freedom, besides dissipative DC transport measurements~\cite{danilenko2023, kringhoj2018, kringhoj2021, sabonis2020}.

\begin{figure*}[t]
\centering
\includegraphics[width=\textwidth]{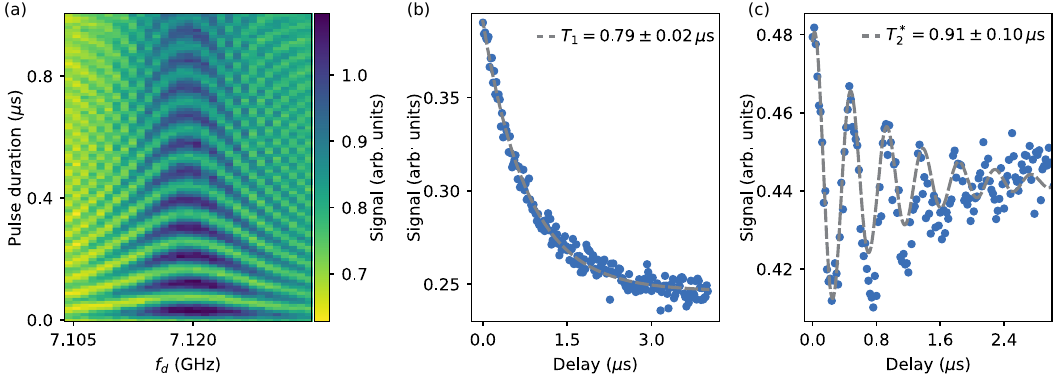}
\caption{Coherent control and time-domain characterization of a shadow-junction gatemon. (a) Rabi chevron measured as a function of drive frequency and pulse duration. (b) Representative energy-relaxation measurement. The solid line is a fit to $A\exp(-\tau/T_1)+B$, yielding $T_1=0.79~\mu\mathrm{s}$. (c) Ramsey interference measured as a function of the delay between two $\pi/2$ pulses. The solid line is a fit to an exponentially damped oscillation, yielding $T_2^*=0.91~\mu\mathrm{s}$.}
\label{fig:coherent_control}
\end{figure*}

Ternary InAs$_\mathrm{1-x}$Sb$_\mathrm{x}$ nanowires offer an unique materials platform in which the semiconductor properties can be engineered through the Sb fraction x~\cite{zhuang2015, sestoft2018, Carrad2022, Stampfer2022}.
At intermidiate compositions, InAs$_\mathrm{x}$Sb$_\mathrm{1-x}$ nanowires have exhibited spin--orbit coupling interaction exceeding that of the bianry InAs and InSb endpoint materials, while retaining large $g$ factor and epitaxial coupling to aluminum (Al) produces a hard induced superconducting gap~\cite{sestoft2018, Carrad2022, Stampfer2022}.
Combining alloy composition, electrostatic confinement, and junction geometry creates a broad parameter space for engineering and investigating hybrid electronic states.
Embedding such a material platform in a cQED architecture enables fast and high-sensitivity microwave measurements of the junction properties.
In the short-junction limit, the Josephson potential of S--Sm--S junction is determined by the transmission probabilities $\{T_i\}$ of the occupied channels,
\begin{equation}
    U(\phi) = -\Delta\sum_i\sqrt{1 - T_i\sin^2{(\phi/2)}},
\end{equation}
where $\Delta$ is the induced superconducting gap and $\phi$ is the superconducting phase difference across the junction~\cite{beenakker1991, kringhoj2018}.
Small changes in Josephson potential produces measureble shifts in the qubit frequencies, while microwave pulse sequences provide access to coherent dynamics, relaxation, dephasing, and temporal fluctuations on nanosecond-to-microsecond time scales.

Here, we demonstrate electrical tuning of the qubit transition over a broad frequency range and coherent control of the qubit states.
We measure energy relaxation time in five gatemon qubits distributed across three chips and compare the measured lifetime with a simpified spontaneous-emission estimate.
Our measurements extend InAsSb/Al junctions from DC transport characterization to coherent cQED operation.

Figure~\ref{fig:design} shows the device architecture and the formation mechanism of the nanowire Josephson junction. 
The gatemon consists of an InAsSb/Al nanowire junction shunted by a large X-shaped capacitor and coupled to a readout resonator, as illustrated in Fig.~\ref{fig:design}(a).
The nanowire connects the superconducting island to ground and is capacitively coupled to a local bottom electrostatic gate through an HfO$_\textrm{x}$ dielectric, as shown in Fig.~\ref{fig:design}(b).
Opposite directional InAsSb nanowires were grown from InAs (111)B trenches using MBE system ~\cite{khan2020, khan2021, khan2025}. After the nanowire growth the substrate was cooled down and Al was deposited within the same growth process. During Al deposition, one nanowire acts as a physical shadow mask for the neighboring nanowire, blocking the incident Al flux over a short segment, as shown in Fig.~\ref{fig:design}(c).
The resulting uncovered InAsSb segment between the two Al-coated regions forms the S--Sm--S Josephson weak link. After growth, the individual nanowire with the shadow-defined junction is transferred from the growth substrate onto the qubit chip via a micro-manipulator.
This shadow-junction technique was previously developed and shown to produce highly transparent Josephson junctions with electrostatically tunable supercurrent~\cite{khan2020, Carrad2022, Stampfer2022}.
Because the weak link is defined during Al deposition, selective removal of the superconducting shell is not required in the active junction region.
Therefore, this technique reduces the post-growth processing applied directly to the weak link and limits its exposure to etchants, residual Al from incomplete shell removal, and possible surface damage~\cite{khan2020, Chen2023, sestoft2024}, as illustrated schematically in Fig.~\ref{fig:design}(d).

We next investigate the microwave response of the InAsSb/Al gatemon.
Figure~\ref{fig:gate_spectroscopy} demonstrates electrostatic control of the readout response and qubit spectrum.
At each gate voltage $V_g$, we first perform resonator spectroscopy to determine the readout resonator frequency, followed by qubit two-tone spectroscopy to identify qubit transition.
Figure~\ref{fig:gate_spectroscopy}(a) and (b) show the resonator and qubit spectra  as a function of $V_g$, respectively.
The data in each panel combine two consecutive gate sweeps acquired over different frequency windows: one from $V_g = 0$ to $-2~\mathrm{V}$ and another from $V_g=-2$ to $-4~\mathrm{V}$.
The two-tone spectroscopy reveals a qubit transition that can be tuned over several gigahertz.
The transition is observed predominantly at comparatively high frequencies.
Within an effective transmon description, this may reflect a relatively large Josephson energy, $E_J=\Phi_0 I_c/(2\pi)$ where $\Phi_0$ is the flux quantum and $I_c$ is the critical current. 
The larger $E_J$ is consistent with the high critical currents expected for transparent shadow-defined InAsSb/Al junctions ~\cite{khan2020, Carrad2022, Stampfer2022}.
The transition exhibits a nonmonotonic dependence on $V_g$, as commonly observed in mesoscopic semiconductor Josephson junctions~\cite{larsen2015, kringhoj2018,feldsteinbofill2025}.
We further characterize the qubit transition at the gate voltage marked by the dashed line in Fig.~\ref{fig:gate_spectroscopy}(b).
Figure~\ref{fig:gate_spectroscopy}(c) shows the two-tone response as a function of drive power and frequency.
From the power-dependent spectrum, we identify the fundamental qubit transition $f_q$ and the two-photon transition $f_{02}/2$.
Using $\alpha = 2(f_{02}/2 - f_q)$, we extract an anharmonicity of $\alpha\approx\SI{-120}{\mega\hertz}$.
The magnitude of the extracted anharmonicity is substaintially smaller than the designed charging energy $E_C\approx\SI{204}{\mega\hertz}$, which sets the approximate anharmonicity of a conventional SIS transmon.
Such reduced anharmonicity has also been observed in other S--Sm--S gatemons and is consistent with a non-sinosoidal current--phase relation arising from highly transmitting transport channels~\cite{kringhoj2018}.

Following the electrostatic control of the qubit transition, we next investigate coherent qubit dynamics using pulsed microwave measurements. 
Figure~\ref{fig:coherent_control} summarizes the Rabi, energy relaxation and Ramsey measurements performed at a selected operating point. Figure~\ref{fig:coherent_control}(a) shows Rabi oscillation as a function of drive frequency $f_d$ and pulse duration $\tau_d$.
The oscillation frequency follows $f_\mathrm{R} = \sqrt{f_{\mathrm{R},0}^2 + (f_d - f_q)^2}$, where $f_{\mathrm{R},0}$ is the resonant Rabi frequency.
The resulting chevron demonstrates coherent qubit control and enables calibration of $\pi$ and $\pi/2$ pulses.
We then measure the energy relaxation time by applying a $\pi$ pulse and varying the delay time before readout.
The representative relaxation trace in Fig.~\ref{fig:coherent_control}(b) is fitted to a exponential decay, yieling $T_1=\SI{0.79}{\micro\second}$.
Finally, we characterize the phase coherence using a Ramsey sequence consisting of two $\pi/2$ pulses separated by a variable decay.
The resulting Ramsey fringes are shown in Fig.~\ref{fig:coherent_control}(c).
Fitting the data to an exponentially decaying cosine yields $T_2^*=\SI{0.91}{\micro\second}$.
Together, the time-domain measurements demonstrate coherent operation of the InAsSb/Al gatemon.

\begin{figure}
\centering
\includegraphics[width=\columnwidth]{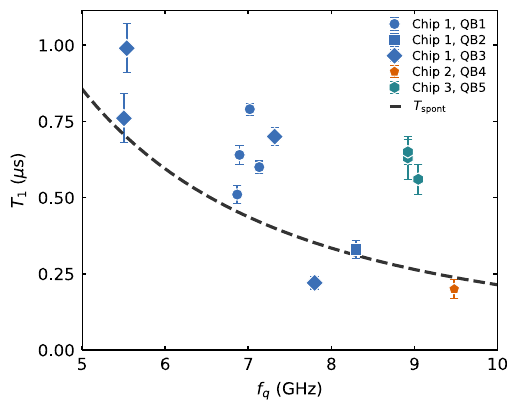}
\caption{Energy-relaxation times of five shadow-junction gatemon qubits distributed across three chips. Each point shows the fitted relaxation time $T_1$ at a selected qubit frequency $f_q$, with error bars indicating the standard uncertainty obtained from the exponential fit. Colors identify the chips and marker shapes distinguish individual qubits. The dashed curve shows the spontaneous-emission lifetime $T_{\mathrm{spont}}$ estimated for a qubit capacitance $C_q=95~\mathrm{fF}$ coupled through $C_c=1.5~\mathrm{fF}$ to a frequency-independent $50~\Omega$ environment. Because the effective microwave impedance and coupling capacitance are not independently determined, the curve represents a simplified circuit estimate rather than a strict upper bound.}
\label{fig:T1_summary}
\end{figure}

To assess device-to-device reproducibility, we measure $T_1$ for five gatemons distributed across three chips.
Figure~\ref{fig:T1_summary} shows the relaxation times between approximately \SI{0.2}{\micro\second} and \SI{1.0}{\micro\second} over the measured qubit-frequency range.
The data exhibit an overall decrease in $T_1$ with increasing qubit frequency.
The dashed curve shows the spontaneous emission time $T_{1, \mathrm{spont}}$ estimated by coupling qubit capacitance $C_q=\SI{95}{\femto\farad}$ through a coupling capacitor $C_c=\SI{1.5}{\femto\farad}$ to a frequency-independent \SI{50}{\ohm} environment~\cite{houck2008}.
The capacitance values are extracted from an Ansys electromagnetic simulation.
Its magnitude is comparable to many of the measured values, indicating that spontaneous emission may contribute to relaxation in the present circuit.
Because the calculation assumes a frequency-independent \SI{50}{\ohm} environment, the dashed curve should be interpreted as a circuit estimate rather than a strict upper bound on $T_1$.
Spontaneous emission is a circuit-level loss channel rather than an intrinsic limitaion of the semiconductor Josephson junction.
Previous work~\cite{sun2026, feldsteinbofill2025} has demonstrated that spontaneous emission can be strongly suppressed through appropriate circuit design and microwave engineering in gatemons.
In the present devices, the close proximity of the gate line to the qubit capacitor results in a comparatively large coupling capacitance to the control environment.
The measured $T_1$ values therefore characterize the complete circuit architecture and should not be intepreted as the intrinsic coherence limit of the shadow-defined InAsSb/Al junction.

In conclusion, we have realized and coherently operated gatemons based on shadow-defined ternary InAsSb/Al nanowire Josephson junctions.
Gate-dependent spectroscopy demonstrates broad electrical tunability, while the power spectroscopy reveals a reduced anharmonicity consistent with a high-transparency semiconductor weak link.
Rabi oscillations and Ramsey interference reveal coherent control.
Across five gatemons distributed over three chips, we measure relaxation times at around \SI{1}{\micro\second}, with an overall decrease toward higher qubit frequencies.
The comparable spontaneous-emission estimate suggests that coupling to the electromagnetic environment contributes to relaxation in the present circuit design.
These measurements establish shadow-defined InAsSb/Al junctions as electrostatically tunable nonlinear elements for coherent cQED circuits and motivate future designs aimed at suppressing circuit-mediated loss and probing mesoscopic properties of the ternary weak link.

\FloatBarrier
\clearpage 
\appendix 
\section{Repeated energy-relaxation measurements}
\label{app:repeated_T1}

\begin{figure}[H]
    \centering 
    \includegraphics[width=\columnwidth ]{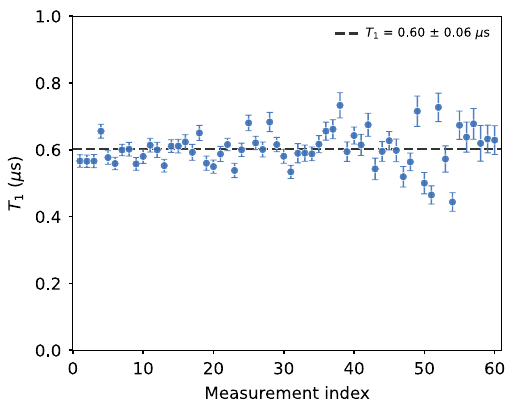} 
    \caption{ Repeated energy-relaxation measurements of an InAsSb/Al gatemon. Each point shows the relaxation time $T_1$ extracted from an independent exponential-decay measurement, with error bars indicating the uncertainty obtained from the fit. The dashed line indicates the mean relaxation time. The 60 measurements yield $\langle T_1\rangle=\SI{0.60(6)}{\micro\second}$, where the uncertainty denotes the standard deviation of the measured distribution. } 
    \label{fig:repeated_T1} 
\end{figure}

To characterize the temporal variation of the energy-relaxation time, we repeat the $T_1$ measurement 60 times at the same operating point. 
Each relaxation trace is fitted to an exponential-decay model. 
The extracted relaxation times are shown in Fig.~\ref{fig:repeated_T1}. 
We obtain a mean relaxation time and standard deviation of $\langle T_1\rangle=\SI{0.60(6)}{\micro\second}$. 

\section{Fridge wiring diagram}
\label{app:fridge_wiring}

\begin{figure}[h]
    \centering 
    \includegraphics[width=\columnwidth ]{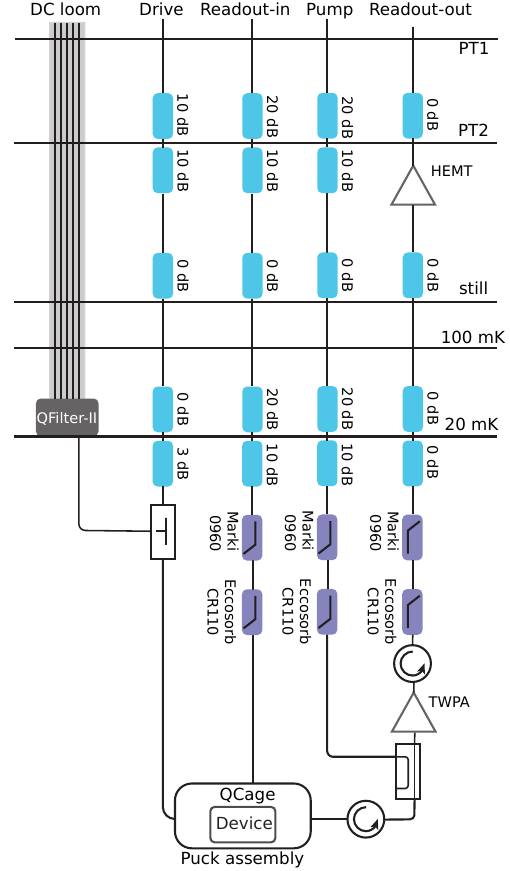} 
    \caption{Dilution fridge wiring diagram. The qubit drive, readout-in, and TWPA-pump lines are attenuated and thermalized at multiple temperature stages. The DC gate signal is filtered by a low-pass filter and combined with the drive signal through a bias-Tee at the mixing chamber stage. The transmitted readout signal is amplified by a traveling-wave parametric amplifier (TWPA) and a high electron mobility transistor (HEMT) amplifier. The device is mounted in a Qcage at the mixing chamber stage.}
    \label{fig:fridge_wiring} 
\end{figure}

The divices were measured in a dilution fridge with a base temperature of around \SI{22}{\milli\kelvin}.
The wiring diagram is shown in Fig.~\ref{fig:fridge_wiring}.

\section{Device fabrication}

The devices were fabricated on high-resistivity silicon substrates coated with a \SI{100}{\nano\meter}-thick Al film deposited inside the MBE system.
The superconducting groud plane, the qubit capacitors, readout resonators, feedline,  and control structures were patterned by electron-beam lithography using PMMA A4 resist.
After exposure, the resist was developed in a MIBK:IPA 1:3 solution.
and the Al film was etched by Transene Aluminum Etchant type D solution.
A second e-beam lithography step was used to define the window for the local gate dielectric.
An approximately \SI{15}{\nano\meter}-thick HfO$_\mathrm{x}$ layer was subsequently deposited by atomic layer deposition (ALD).
The nanowires were determistically transferred from the growth substrate onto the qubit chip via a micro-manipulator.
Atomic force microscopy (AFM) and fast scanning electron microscopy (SEM) were performed to identify if the nanowires containing shadowe-defined junctions and to verify the alignment of each junction with its corresponding bottom gate.
For electrical contacting, the devices were coated with a trilayer resist stack consisting of two MMA EL9 layers and one PMMA A4 layer.
After e-beam exposure and development of the contact pattern, in situ Ar ion milling was performed to remove the native oxide from the contact regions.
A \SI{400}{\nano\meter}-thick Al layer was then deposited to form galvanic contacts between the nanowires and the superconducting microwave circuit.

\begin{acknowledgments}
The authors gratefully acknowledge Smitha Nair for assistance with device fabrication, Peter Krogstrup for supervising the nanowire growth, and Andr\'as Gyenis for the fruitful discussions.
This research was supported by the Novo Nordisk Foundation (grant no. NNF22SA0081175), the NNF Quantum Computing Programme, the Villum Foundation through a Villum Young Investigator grant (grant no. 37467), Innovation Fund Denmark (grant no. 2081-00013B, DanQ), the U.S. Army Research Office (grant no. W911NF-22-1-0042, NHyDTech), the European Union through an ERC Starting Grant (grant no. 101077479, NovADePro), and the Carlsberg Foundation (grant no. CF21-0343). SAK acknowledges funding from Danish Agency for Higher Education and Science, DIANA Quantum Center (DIANAQ), The Danish Chips Competence Center (DKCCC), DIRAC: A national CERN lighthouse initiative, Innovation Fund Denmark (IFD) projects: DeQD, QLIGHT, European Metrology projects: MetSuperQ (Project No. 23FUN08), NoQTeS (Project No. 23NRM04), DINAMO (Project No. 24DIT02). Any opinions, findings, conclusions, or recommendations expressed in this material are those of the authors and do not necessarily reflect the views of the U.S. Army Research Office or the U.S. Government. Views and opinions expressed are those of the authors only and do not necessarily reflect those of the European Union or the European Research Council. Neither the European Union nor the granting authority can be held responsible for them. We thank Lena Jacobsen for program-management support.
\end{acknowledgments}

\end{document}